\documentclass[useAMS,usenatbib]{mn2e}

\usepackage{graphicx,amssymb,bm}
\usepackage{subfig,xfrac}
\usepackage{float, Abbrev}
\usepackage[fleqn]{amsmath}  
\usepackage{color}
\usepackage[draft]{hyperref}

\usepackage{multirow}

\newcommand{\cmg}{cm$^2$/g}

\newcommand{\be}{\begin{equation}}
\newcommand{\ee}{\end{equation}}
\newcommand{\ba}{\begin{eqnarray}}
\newcommand{\ea}{\end{eqnarray}}

\newcommand{\sdm}{\sigma_{\rm DM}/m}

\def\simlt{\lower.5ex\hbox{$\; \buildrel < \over \sim \;$}}

\newcommand{\fig}{\begin{figure} \begin{center}}
\newcommand{\efig}{\end{center}\end{figure} }
\newcommand{\figs}{\begin{figure*}\begin{minipage}{180mm} \begin{center}}
\newcommand{\efigs}{\end{center}\end{minipage}\end{figure*} }
\def\simgt{\lower.5ex\hbox{$\; \buildrel > \over \sim \;$}}

\usepackage{graphicx} 

\title[IA with SIDM]{The impact of Self-Interacting Dark Matter on the Intrinsic Alignments of Galaxies}
\author[D. Harvey et al]
{David Harvey$^{1}$\thanks{e-mail: {\tt harvey@lorentz.leidenuniv.nl}}, {Nora Elisa Chisari}$^2$ and {Andrew Robertson}$^3$ \\
$^{1}$Lorentz Institute, Leiden University, Niels Bohrweg 2, Leiden, NL-2333 CA, The Netherlands.
\\
$^{2}$Institute for Theoretical Physics, Utrecht University, Princetonplein 5, 3584 CC, The Netherlands.
\\
$^{3}$ Institute for Computational Cosmology, Department of Physics, Durham University, South Road, Durham DH1 3LE, UK}
\begin{document}

\date{Accepted ---. Received ---; in original form \today.}

\pagerange{\pageref{firstpage}--\pageref{lastpage}} \pubyear{2017}

\maketitle

\label{firstpage}

\begin{abstract}
The formation and evolution of galaxies is known to be sensitive to tidal processes leading to intrinsic correlations between their shapes and orientations. Such correlations can be measured to high significance today, suggesting that cosmological information can be extracted from them. Among the most pressing questions in particle physics and cosmology is the nature of dark matter. If dark matter is self-interacting, it can leave an imprint on galaxy shapes. In this work, we investigate whether self-interactions can produce a long-lasting imprint on intrinsic galaxy shape correlations. We investigate this observable at low redshift ($z<0.4$) using a state-of-the-art suite of cosmological hydro-dynamical simulations where the dark matter model is varied. We find that dark matter self-interactions induce a mass dependent suppression in the intrinsic alignment signal by up to 50\% out to tens of mega-parsecs, showing that self-interactions can impact structure outside the very core of clusters. We find evidence that self-interactions have a scale-dependent impact on the intrinsic alignment signal that is sufficiently different from signatures introduced by differing baryonic physics prescriptions, suggesting that it is detectable with up-coming all-sky surveys.
\end{abstract}


\section{Introduction}

Galaxy shapes, both the magnitude of their ellipticity and position angles, are coherently aligned by large-scale tidal fields, sourced by over-densities in the Universe \citep{CroftMetzler2000,LeePen2000,Heavens2000,Catelan2001,Crittenden2001,Mackey2002,Heymans2004,Aubert2004}. Confirmed observationally \citep{Brown2002,Mandelbaum2006,Hirata2007,Joachimi2011,Heymans2006,Kirk2012,Okumura2009,Singh2015,SinghMandelbaum2016,kidsIA}, the ``intrinsic alignments" (IA) of galaxies with over-densities are a serious contaminant of weak lensing cosmic shear studies, introducing additional correlations and potentially biasing the inference of cosmological parameters \citep{Hirata2004,Bridle2007,Kirk2010, Kirk2012,Krause2016}. Frameworks to mitigate the impact of IA on cosmological model inference have been developed, whereby nuisance parameters can be marginalised over \citep{King2005,Joachimi2011}. This requires insights into the origin of the IA signal, which has resulted in several efforts to analytically model the signal from physical first principles \citep{Catelan2001}. Initial models assumed that the over-densities grow linearly, whereas in practice this is only true down to separations of $\sim 10$ $h^{-1}$ Mpc. More recent efforts have extended this modelling to the quasi-linear regime via perturbative expansions \citep{Blazek2011,Blazek2015,Blazek2019} or effective field theory approaches \citep{Vlah2020,Vlah20}, and to the one halo regime via the halo model \citep{SchneiderBridle,Fortuna}. 
In parallel, progress in hydro-dynamical cosmological simulations, which include complex baryonic physics processes such as feedback from supernova and active galactic nuclei (AGN), has enabled predictions of the IA signal down to $\sim0.1$ $h^{-1}$ Mpc and comparison with observations \citep{Codis2015,Tenneti2015,Velliscig2015,Chisari2015,Chisari2016,Hilbert2017,Kraljic20,Samuroff20,Shi20}. These simulations help validate and calibrate analytical methods, whilst providing some priors and physical insight into the observed correlations. 

While most efforts remain focused on mitigation of IA, they themselves are rich with cosmological information. In particular, alignments are sensitive to Baryon Acoustic Oscillations \citep{Chisari2013}, an-isotropic primordial non-Gaussianity \citep{Schmidt2015,Chisari2016}, primordial gravitational waves \citep{Chisari2014b,Biagetti2020} and modified gravity \citep{Huillier2017}. Different strategies can be useful for extracting this information, including combining multiple shape measurements \citep{SinghMandelbaum2016,Chisari2016b} or exploiting the mass-dependence of the IA signal \citep{Piras2018}. In this paper, we investigate how altering the interacting properties of dark matter may influence the intrinsic alignment of galaxies with over-densities in the Universe. 

Despite the success of the cold and collisionless dark matter paradigm (CDM), we know very little about the particle nature of dark matter, other than its interactions with the Standard Model (protons, neutrons, neutrinos, etc.) must be exceptionally weak. Interestingly, the self-interaction of dark matter is not constrained by the same limits and provides a unique avenue to understand forces within the dark sector without having to assume any coupling to the Standard Model (for a review see \cite{SIDM_review}). It is thus crucial to constrain the self-interacting dark matter (SIDM) cross-section, most often normalized by its mass: $\sigma_{\rm DM}/m$. Such studies often concentrate on the inner profiles of either high or low mass halos \citep{cuspDraco,Harvey_dwarf,densityProf2,Harvey_BCG,Bondarenko2020}. Recently, it has been suggested that the anti-correlation between the peri-centre of the dwarf galaxies and their dark matter central density can be only explained by SIDM \citep{Kaplinghat2019}, resulting in a velocity-dependent cross-section (vdSIDM) with $\sigma_{\rm DM}/m\sim100$cm$^2$/g at the Dwarf Scale, and $\sigma_{\rm DM}/m\sim0.1$ cm$^2$/g at cluster scale \citep{Correa2020}. Indeed, this would be consistent with cluster scale constraints that limit the cross-section to $\sigma_{\rm DM}< 0.5$ cm$^2$/g \citep{SIDM_BAHAMAS,Sagunski}.  

In this work, we rely on the adapted BAHAMAS-SIDM suite of cosmological simulations to carry out a study of large-scale signatures of SIDM as probed by galaxies around clusters of galaxies in a model where scatterings are late-time and elastic (without any change to the standard $\Lambda$CDM primordial power spectrum). Due to the limited mass resolution of the simulations, and because alignments are stronger for higher mass galaxies \citep{Piras2018}, we restrict our study to central galaxies alone. This ensures that  we have a complete sample of well resolved galaxies. We specifically demonstrate that the intrinsic alignments of galaxies are sensitive to the nature of dark matter up to scales of several $h^{-1} $Mpc. 

The manuscript is set out as follows: in section \ref{sec:sims} we outline the cosmological simulations used, in section \ref{sec:corr} we set out the correlation functions, section \ref{sec:results} shows our results, in section \ref{sec:disc} we discuss our results and in section \ref{sec:conclusions} we conclude. Throughout this paper we assume a WMAP9 cosmology \citep{WMAP9cosmo} with $\Omega_{\rm M} = 0.2793$, $\Omega_{\rm B}  = 0.0463$, $\Omega_{\rm \Lambda}  0.7207$, $\sigma_{\rm 8} = 0.812$, $ns = 0.972$ and $h = 0.700$.

\section{Cosmological simulations of Self-Interacting Dark Matter}
\label{sec:sims}

The suite of simulations employed here (BAHAMAS-SIDM) are those laid out in \cite{RobertsonBAHAMAS} and we briefly summarise their properties here.  Specifically, BAHAMAS-SIDM was produced by including an implementation of dark matter scattering (SIDM)  within the BAryons And HAloes of MAssive Systems (BAHAMAS) model of \citet{BAHAMAS,BAHAMASB}.  The original BAHAMAS suite of simulations includes a number of cosmological hydrodynamical simulations run with the TreePM-SPH code {\sc GADGET3} \citep{Springel05}.  BAHAMAS includes subgrid physics originally developed for the OWLS project \citep{OWLS} for processes that are not directly resolved in the simulations, including radiative-cooling \citep{OWLScoolingRates}, stellar evolution and chemodynamics \citep{OWLSstellarEvoChemo}, star formation \citep{OWLSstarFormation}, and stellar and AGN feedback \citep{OWLSagn,OWLSagn2}.

The BAHAMAS runs accurately reproduce the local galaxy stellar mass function and the gas mass fractions of galaxy groups/clusters, a result of calibration of the AGN and stellar feedback parameters.  However, as shown in \citet{BAHAMAS}, the simulations also reproduce a wide range of other observables (e.g., X-ray and tSZ scaling relations, hot gas profiles, evolution of the GSMF, QSO luminosity functions, etc.) without any explicit calibration to do so. 

For BAHAMAS-SIDM, the parameters dictating the efficiencies of stellar and AGN feedback were left unchanged from their calibrated values in BAHAMAS.  Furthermore, like BAHAMAS, the BAHAMAS-SIDM runs of \citet{RobertsonBAHAMAS} were carried out in large periodic boxes whose side measures $400$ $h^{-1}$ Mpc, and they adopt a Plummer-equivalent softening length of $4$ $h^{-1}$ kpc (physical) below $z=3$.  Given that the aim of this study is to measure the impact of SIDM on the intrinsic alignments of galaxies out to tens of mega-parsecs, the BAHAMAS simulations provides the ideal volume and resolution to garner sufficient statistics.

\begin{figure*}
\includegraphics[width=0.89\textwidth]{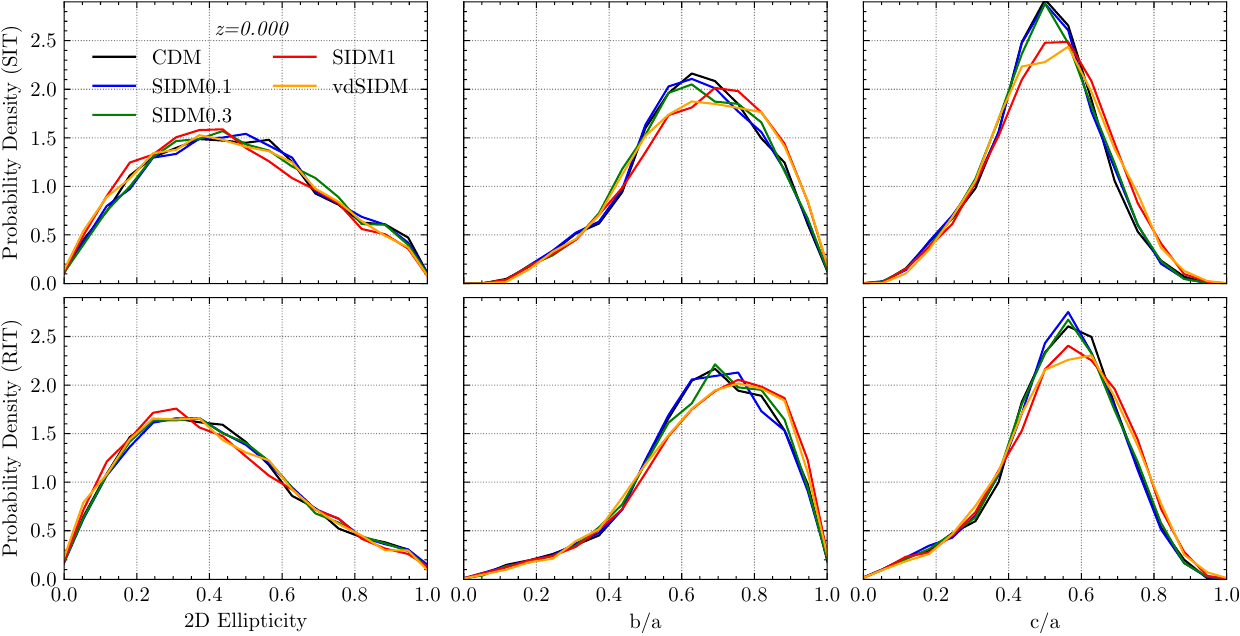}
\caption{\label{fig:shape_distribution} The total distribution of 2D \& 3D shapes integrated for $z=0$, for the simple inertia tensor (SIT, top row) and the reduced inertia tensor (RIT, bottom row). Each colour represents a different dark matter model where we show CDM (black), $\sdm=0.1$\cmg (blue), $\sdm=0.3$\cmg (green), $\sdm=1.0$\cmg (red) and vdSIDM (yellow). All shapes are measured within the fiducial radius of 100 physical kpc and using Eq. (\ref{eq:it}).}
\end{figure*}

We obtain IA predictions from the various SIDM models laid out in \cite{RobertsonBAHAMAS} that include (in addition to CDM) three velocity-{\it independent} cross-sections of $\sigma_{\rm DM}/m=0.1, 0.3, 1.0$ cm$^2$/g and one velocity-{\it dependent} cross-section, where the differential cross-section is defined as
\be
\frac{d\sigma}{d\Omega} = \frac{\sigma_0}{4\pi\left( 1 + \left(\frac{v}{w}\right)^2\sin^2(\frac{\theta}{2})\right)^2}.
\ee
Here, $v$ is the relative velocity of two interacting particles, $\sigma_0$ and $w$ are parameters of the model corresponding to the cross-section at low-velocities (where it becomes velocity independent), and the turn-over velocity. Following this, the momentum transfer cross-section is given by the integral over all solid angles, $\Omega$. 

 We simulate a velocity-{\it dependent} cross-section with a normalisation of $\sigma_0=3.04$ cm$^2$/g and a turn-over velocity of $w=560$ km/s, corresponding to the best fitting model from \cite{Kaplinghat2016} (corresponding to $\sigma_{\rm DM}\sim1$cm${^2}$/g on cluster scales). BAHAMAS-SIDM adopts the same parameters for the sub-grid modelling as the original BAHAMAS runs, since they remain unchanged within the sensitivity of the observations (see Discussion for more) \citep{RobertsonBAHAMAS}. These cross-sections span a regime which includes both already excluded and interesting SIDM models. Constraints have limited the cross-section to $\sigma_{\rm DM}\lesssim0.5$ cm${^2}$/g \citep{SIDM_BAHAMAS,Sagunski}, with some observations suggesting that dark matter could self-interact at $\sigma_{\rm DM}\sim0.1$cm${^2}$/g on cluster scales \citep{Correa2020}. 

We extract snapshots from four redshift slices: $z=\{0., 0.125, 0.250, 0.375\}$. We focus on this redshift range motivated by significant detections of the alignment signal in low redshift observations \citep[e.g.][]{Singh2015}. We identify haloes and sub-haloes within the simulation box using the {\sc SUBFIND} algorithm \citep{springel01} and extract all galaxies that have a stellar mass $M^*>10^{10}M_\odot/h$. The entire sample of galaxies is referred to as the ``D'' sample. 

Intrinsic alignments are known to be mass-dependent, with higher mass galaxies exhibiting stronger alignment trends \citep{vanUitert17,Piras2018}. We hence hypothesise that, if an impact from dark matter self-interactions is present in the alignment statistics, it will be more prominent for the shapes of centrals. This is indeed verified in our results, and we refer to the sample of centrals from the simulations in what follows as the ``S'' sample. Thus, we focus on describing the alignments of the galaxies in S and around density tracers D for the rest of the manuscript. Notice that due to the limitations imposed by a fixed mass resolution, the galaxy samples vary slightly from one simulation run to the other. This is caused by the suppression of the mass function as a result of the self-interactions.

\subsection{Shape Measurement}

We calculate the shape of a galaxy via its inertia tensor,
\be
I_{ij} = \frac{ \sum_n x_{i,n}x_{j,n}w_n}{\sum w_n},\label{eq:it}
\ee
where $(x_{i,n}, x_{j,n})$ are the coordinates of the particle and $w_n$ is the weight of particle $n$. Intrinsic alignments of low redshift galaxies have been shown to be sensitive to the shape measurement method both in simulations \citep{Chisari2015,Velliscig2015,Tenneti2015} and in observations \citep{SinghMandelbaum2016,Georgiou2019a,Georgiou2019b}. There is evidence for the outskirts of galaxies to be more aligned with each other than their inner regions, a fact that can potentially be exploited for cosmological purposes \citep{Chisari2016b}. To mimic such an effect, we adopt two different weighting schemes in Eq. (\ref{eq:it}): 
simple (SIT) or reduced (RIT). For the SIT, $w_n$ is simply the mass of the particle, $m_n$ (i.e $w_n=m_n$), and for the RIT, it includes the inverse square of the projected distance the particle is from the centre of the halo, (i.e. $w_n=m_n/r_n^2$). 

To measure the shapes of the central galaxies we carry out an iterative process whereby we calculate the moment of inertia of all particles within some given radius of the galaxy centre (and hence the ellipticity). We then re-calculate the moment inertia except now all particles within an elliptical radius defined by the previous iteration estimate of the shape, keeping the area of the ellipse constant (defined by the radius of the initial circle). We continue this iterative process until the shape estimated by two consecutive iterations are within $1\%$ of one another. We initialise this process assuming an ellipticity of zero. 
Throughout the manuscript we measure the shape of each galaxy within a physical distance of $100$ kpc, unless otherwise stated.
However, based on the observational evidence for scale-dependence of IA measurements quoted above, we also perform a second measurement with SIT shapes within 30 physical kpc. We have verified that the alignments of galaxies in sample S are indeed weaker if measured at this scale.

We calculate two different inertia tensors: one three-dimensional tensor and a projected two-dimensional tensor. The projected inertia tensor can be more directly compared to observed galaxy shapes in photometric surveys, but accessing the three-dimensional information from the simulation allows us to investigate the impact of SIDM in IA to higher significance. We then denote the eigenvalues of each tensor as $a^2$, $b^2$ (and $c^2$), where $a^2>b^2>c^2$, each with a corresponding eigenvector defining its direction. We calculate the position angle of the major-axis, $a$ from its eigenvector and denote this $\theta$. The two-dimensional ellipticity of a galaxy is defined as 
\be
e_{+,\times} = \frac{a^2-b^2}{a^2+b^2}[\cos(2\theta), \sin(2\theta)].\label{eq:2dshape2}
\ee
The total 2D ellipticity of the galaxy is thus defined as $e=(a^2-b^2)/(a^2+b^2)$.

Figure \ref{fig:shape_distribution} shows the distribution of projected 2D ellipticities (left hand column), and 3D shapes with the ratios between the second-largest axis and the largest axis (semi-major, middle column) and smallest axis (semi-minor) and the semi-major axis (right hand column) for the four different dark matter models (varying colours). The top row shows the SIT and the bottom row, the RIT. All the curves in this plot correspond to shape measurements taken within 100 physical kpc. We refer results for shapes starting from 30 physical kpc in Section \ref{sec:30}. We note that the distributions may seem immediately broader than what is observed in the literature, however this is due to the choice of ellipticity in equation \eqref{eq:2dshape2}. Often $|e| = (a-b)/(a+b)$ is chosen, resulting in a narrower distribution.

RIT shapes result in higher values for the 3D axis ratios. This is expected, since the RIT measurement tends to circularize them. A similar effect can be seen in the 2D shapes, where RIT results in lower values of $e$. An increasing value of SIDM cross-section results in further circularization of the shapes. This results agrees with previous studies looking at the shape of dark matter halos in SIDM cosmologies \citep{SIDM_shapes,RobertsonBAHAMAS}. The velocity-dependent model yields shapes with a distribution that is very similar to the SIDM1 model, where $\sigma_{\rm DM}=1.0$ cm$^2/$g. This is expected since the effective mass of the sample is $\sim10^{14}M_\odot$ and at this halo mass $\sigma_{\rm DM}(v=500$km/s$)\sim1.0$ cm$^2/$g \citep{RobertsonBAHAMAS}. 

A priori, it is impossible to predict the impact of SIDM on IA from Figure \ref{fig:shape_distribution} alone. In principle, rounder shapes could lead to either increasing the noise in the alignment signal or to suppressing it, or both. Section \ref{sec:corr} introduces the estimators we adopt for investigating the degree of {\it correlated} impact of SIDM on central galaxy alignments with the large-scale structure. 

\fig
\includegraphics[width=0.5\textwidth]{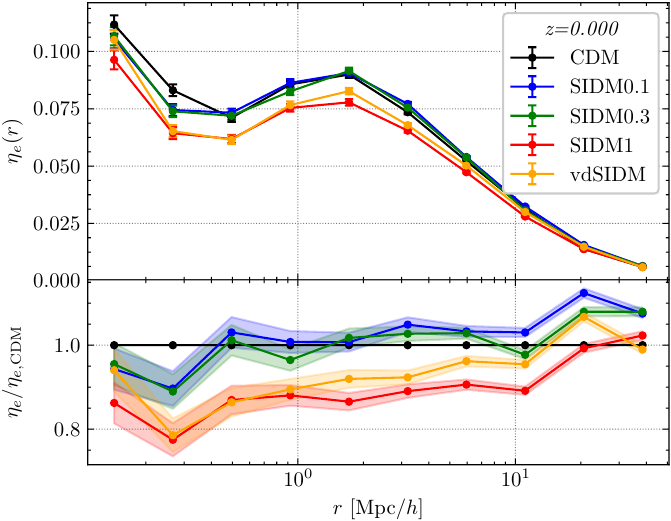}
\caption{\label{fig:3d_correlation} Three-dimensional correlation between the direction of major axes of central galaxies and the separation vector towards all other galaxies at $z=0$. {\it Top:} We show the absolute three-dimensional correlation (equation \ref{eqn:3dcorr}) for each dark matter model. {\it Bottom:} The ratio of the correlation in each model to the CDM case.}
\efig

\section{Correlation Functions}\label{sec:corr}

We calculate the correlation between the shape of central galaxies S, with the positions of all density tracers, i.e galaxies D, and the correlation between the position of centrals with the positions of all galaxies. For normalization purposes, we define a sample of randomly distributed points in the simulation box as $R_S$. This has the same abundance as the sample with galaxy shapes. Analogously, randomly distributed density tracers are labelled $R_D$. 

In a cosmological simulation, availability of three-dimensional shapes for galaxies allows one to define a three-dimensional alignment correlation function as
\be \label{eqn:3dcorr}
\eta_e(r) = \langle |\hat{\Vec{r}}\cdot\hat{\Vec{u}}({\Vec{x}+\Vec{r}})|^2\rangle - 1/3,
\ee
where the hats correspond to unit vectors, $u$ is the direction of the major-axis of a galaxy at position $\vec{x}$ and $r$ is the co-moving separation between galaxies, averaged over all galaxy-central pairs with separation $r$. A positive correlation corresponds to the major-axis pointing parallel to the three-dimensional distance vector separating the two points. 

For the projected shapes, we refer to the tangential component of the ellipticity as $S_+$ and to the cross component as $S_\times$ (Eq. \ref{eq:2dshape2}). We define an estimator for the real-space, normalised correlation function of galaxy shapes and density tracers as

\fig
\includegraphics[width=0.49\textwidth]{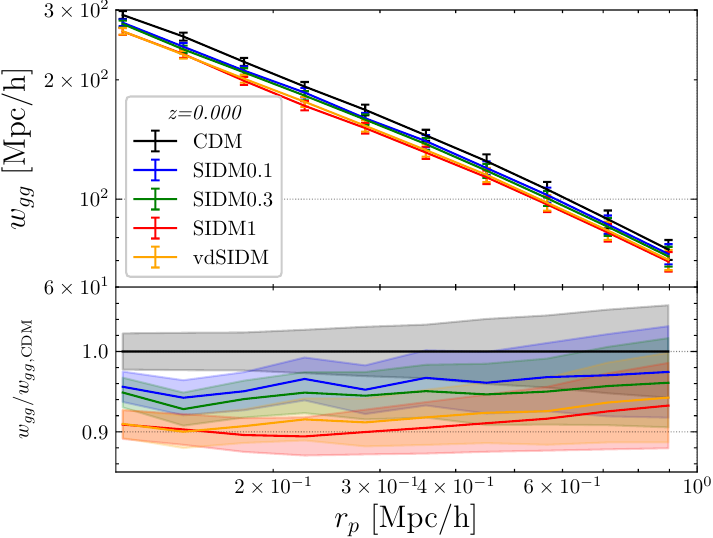}
\caption{\label{fig:totalCorrelations_clustering} Auto-correlation of galaxy clustering for samples D out to $1$Mpc/h. {\it Top: } We show the clustering correlation of all galaxies for different dark matter models. {\it Bottom: } We show the ratio of the projected correlation for each model relative to the CDM model.
 }
\efig

\be
\xi_{g+}(r_p, \Pi) = \frac{S_+ D}{R_SR_D}.
\ee
This is a function of projected co-moving separation, $r_p$, and line-of-sight co-moving separation, $\Pi$, where,
\be
S_+D = \sum_{(r_p,\Pi)} \frac{e_{+,j}}{2\mathcal{R}}.
\ee
Here, $\mathcal{R}$ represents the responsivity of a shape to gravitational shear \citep{shearResponsivity}, $\mathcal{R}=1-\langle e^2\rangle$, where $e_{+,j}$ is the tangential component of the ellipticity for the $j$-th galaxy and $\langle e^2\rangle$ is the root mean square per ellipticity component. For completeness, we include this factor here, although we are only concerned with the scale-dependence of intrinsic alignment correlations. The responsivity factor is only relevant in studies focusing on mitigation of IA in weak lensing cosmology.

\begin{figure*}
\includegraphics[width=0.49\textwidth]{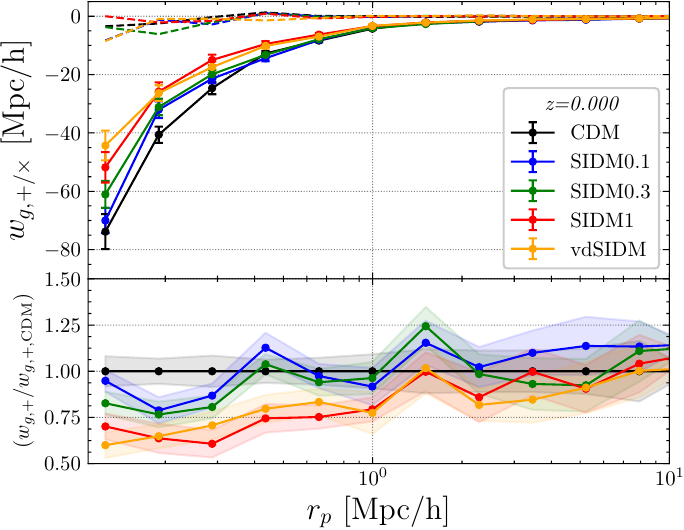}
\includegraphics[width=0.49\textwidth]{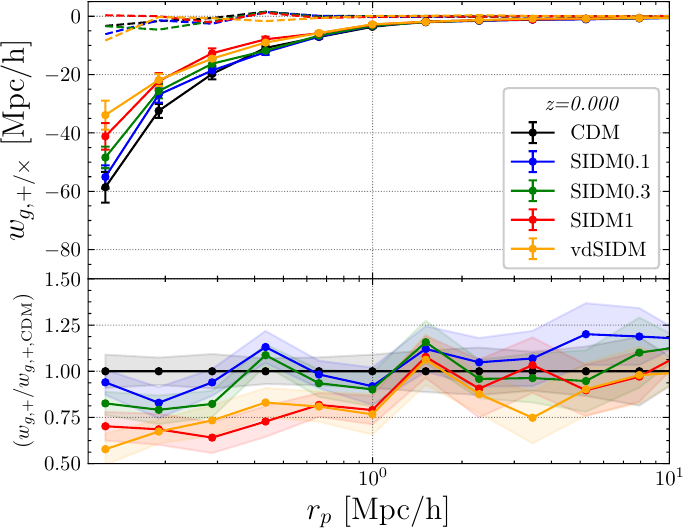}
\caption{\label{fig:totalCorrelations_alignments} 
Projected shape-position correlation between galaxies in sample S and sample D for $z=0$. 
{\it Left:} The absolute projected intrinsic alignment correlation for different dark matter models using SIT shapes measured at $r_{\rm meas}=\,100$kpc and integrated over all redshifts. Both the $w_{g+}$ (solid) and $w_{g\times}$ (dashed) correlations are shown. The bottom panel compares signal relative to CDM. {\it Right:} Analogous results for RIT.
 }
\end{figure*}

We integrate over all line-of-sight bins to get the projected shape-position correlation function, $w_{g+}$,
\be
w_{g+}=\int_{-\Pi_{\rm max}}^{+\Pi_{\rm max}}\,d\Pi\,\xi_{g+}(r_p, \Pi), \label{eq:wg+}
\ee
and similarly for $w_{g\times}$. We choose our projection length to be $\Pi_{\rm max} = 100$ $h^{-1}$ Mpc. 
\cite{Mandelbaum2006} and \cite{Hirata2007} found that integrating beyond $60$ $h^{-1}$ Mpc had no impact on the intrinsic alignment signal. Most of the IA signal is localised in narrow line-of-sight bins due to its physical origin (versus the integrated nature of the gravitational lensing effect).
In order to obtain an estimate of the error in the intrinsic alignment signal, $\sigma_{wg+, {\rm SIDM}}$, we jackknife the sample of galaxies by splitting each volume in to $9$ sub-volumes. This acts to provide us a signal-to-noise estimate for each dark matter model. As such we show the signal to noise relative to a benchmark value (for instance CDM or $z=0$) in each plot, i.e. $(w_{g+, {\rm SIDM}} - w_{g+,{\rm CDM}})/\sigma_{wg+, {\rm SIDM}}$. 

\figs
\includegraphics[width=\textwidth]{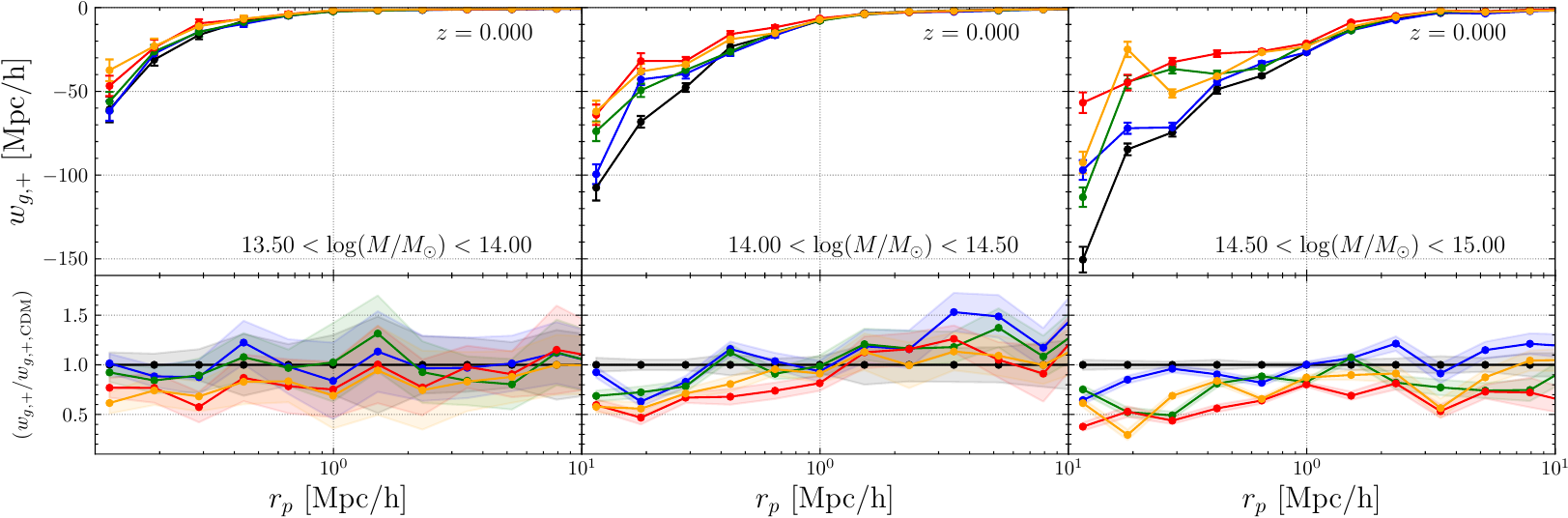}
\caption{\label{fig:IA_mass} The mass dependence of the intrinsic alignment signal for three mass bins (denoted by the legend in each panel). The top panels show the absolute signal for each dark matter model: CDM (black), $\sdm=0.1$\cmg (blue), $\sdm=0.3$\cmg (green), $\sdm=1.0$\cmg (red) and the velocity dependent cross-section ``vdSIDM'' (yellow). The bottom panels show the signal relative to CDM.}
\efigs

\figs
\includegraphics[width=\textwidth]{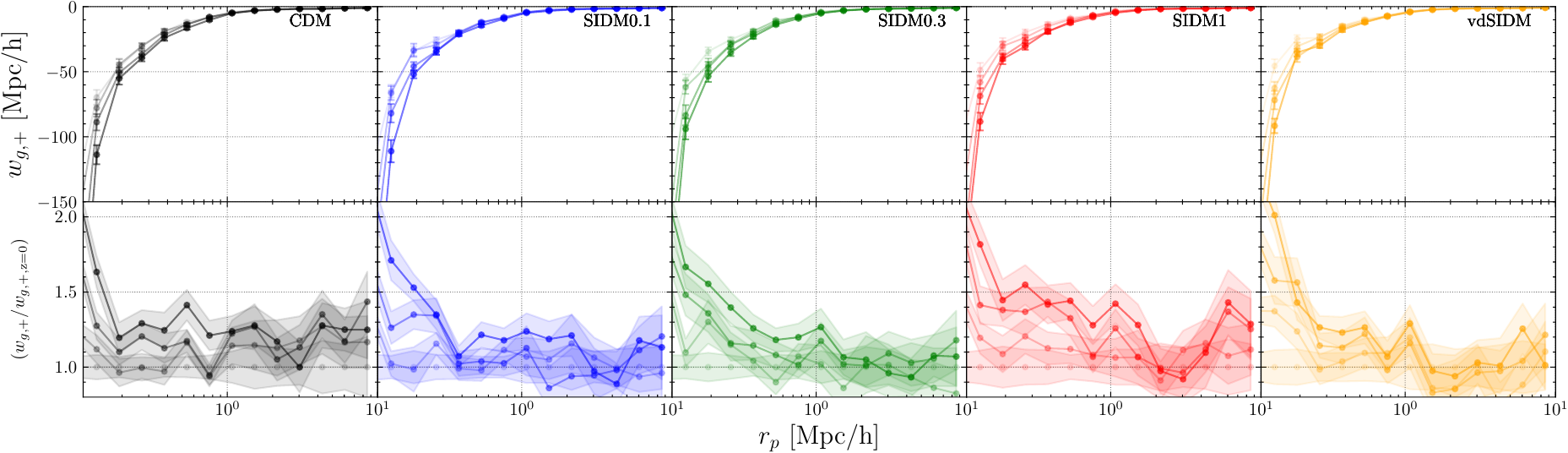}
\caption{\label{fig:IA_redshift} Redshift dependence of each simulation for the 4800 most massive halos in each redshift slice ($z=0,0.125,0.250,0.375$, increasingly solid colour).{\it Top:} The absolute signal for the different dark matter models. {\it Bottom:} We show the signal relative to $z=0$.}
\efigs

Since $w_{g+}$ is sensitive to the clustering bias of galaxies as well as their IA bias, we complement this measurement by also obtaining the clustering (position-position) correlation function for the cross-correlation of S and D. This allows us to determine whether any impact of SIDM observed in $w_{g+}$ might be a consequence of a change in the clustering properties of the samples. The projected clustering correlation, $w_{gg}$, is given by 
\be
w_{gg}=\int_{-\Pi_{\rm max}}^{+\Pi_{\rm max}}\,d\Pi\,\xi_{gg}(r_p, \Pi), \label{eq:wgg}
\ee
where 
\be
\xi_{gg}(r_p, \Pi)=\frac{SD}{R_SR_D}-1.
\ee


Unless explicitly exploring the redshift dependence of our results (discussed in Section \ref{sec:z}), all the figures presented are the results of the redshift stated in the legend of each caption.

\section{Results}\label{sec:results}

\subsection{Sensitivity of IA to SIDM}

We begin by examining the three-dimensional correlation function (c.f. equation \ref{eqn:3dcorr}) to see if self-interactions have an impact on the alignment of central galaxies around density tracers. The top panel of Figure \ref{fig:3d_correlation} shows the three-dimensional alignment signal for the five different dark matter models. We show in the bottom panel of Figure \ref{fig:3d_correlation} the alignment signal in the interacting dark matter models relative to the CDM value. Self-interactions yield a significant suppression in the IA signal that persists to large scales. This suppression is more prominent for increasing cross-sections and can be observed between central-galaxy pairs separated by distances greater than 100 kpc.

These results are consistent with the circularisation of the shapes produced by SIDM in Figure \ref{fig:shape_distribution}. Such a process would introduce random noise in the orientation of the galactic major axes and suppress the alignment correlation. This frequently used statistic only takes into account the galaxy orientations, without considering whether the dark matter model can induce correlated changes in galaxy axis ratios. This is remedied when considering the projected shape correlation of Eq. (\ref{eq:wg+}).

\figs
\includegraphics[width=\textwidth]{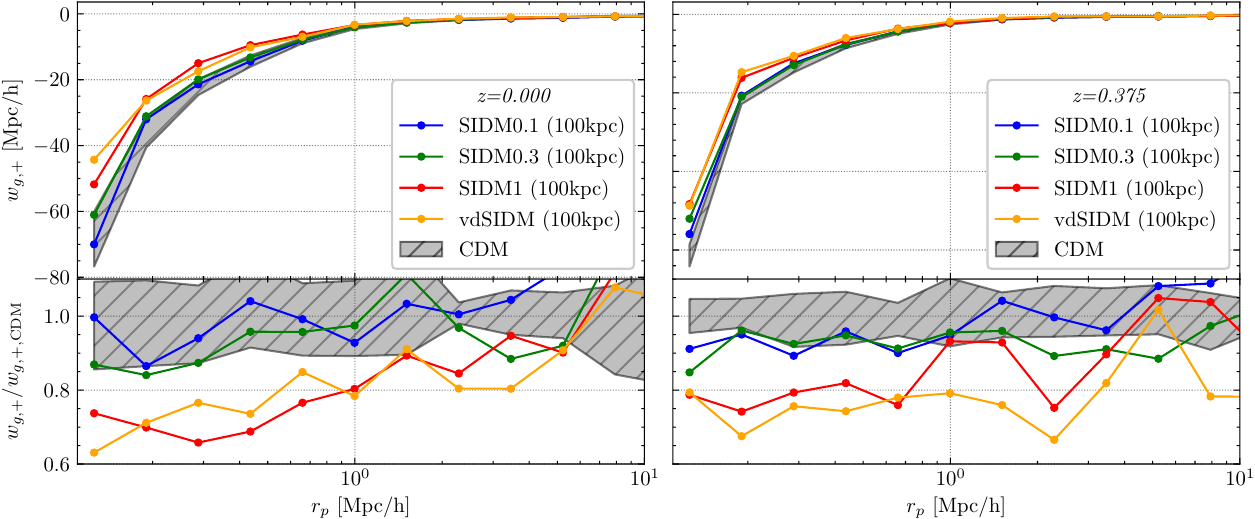} 
\caption{\label{fig:agnDependence} Impact of baryonic feedback on projected alignment statistics for two redshift slices (left, $z=0$ and right, $z=0.375$). {\it Top:} The shaded region shows the expected uncertainty between two extreme AGN models. We show SIDM0.1 (blue) and SIDM1 (red) for a reference. {\it Bottom:} Comparison of the uncertainty in each dark matter model with the uncertainty in the AGN propagated through.}
\efigs
\fig
\includegraphics[width=0.49\textwidth]{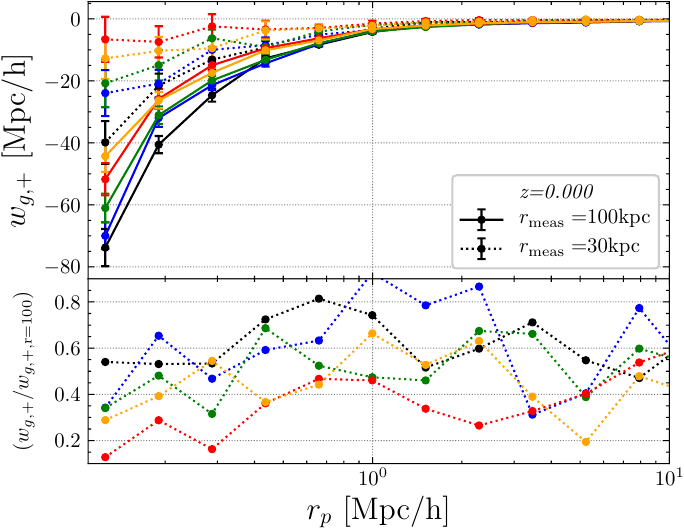}
\caption{\label{fig:scaleDep} The top panel shows the alignment signal for two different shape measurements: within 30 kpc (dotted) and 100 kpc (solid). As expected, the alignment amplitude is smaller when shapes are measured in the inner region. The bottom panel shows the ratio between the alignment signal with the two shape measurements. This ratio depends on the dark matter interaction model, with more interacting model showing increased differences in alignments with scale.}
\efig

\begin{figure}
\includegraphics[width=0.45\textwidth]{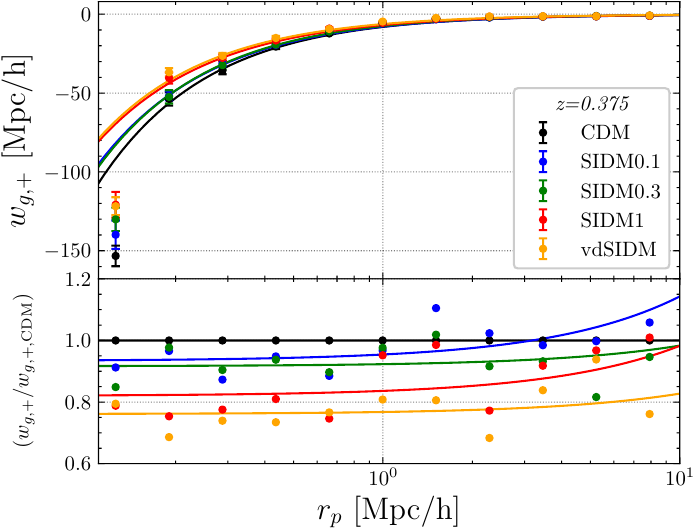}
\includegraphics[width=0.45\textwidth]{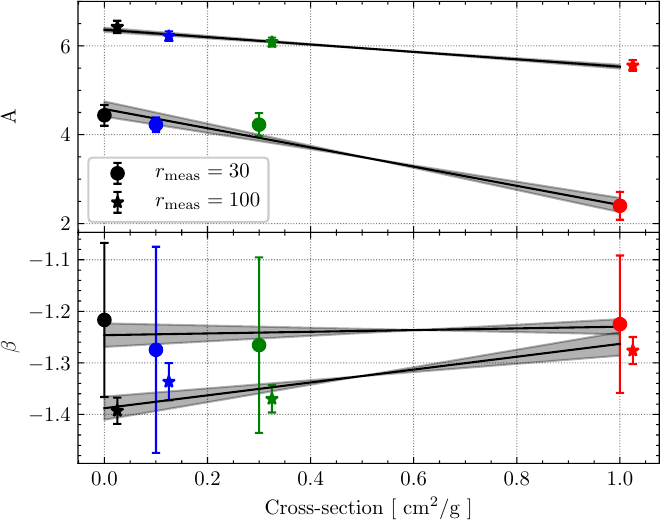}
\includegraphics[width=0.45\textwidth]{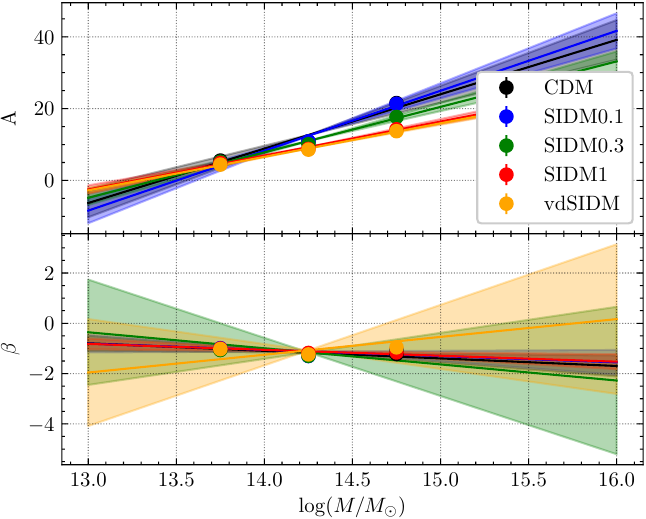}
\caption{\label{fig:IAamplitude} Estimates of the amplitude, $A$ and power law index, $\beta$ fitted to each IA signal. {\it From top to bottom:} The first panel shows the direct power-law fits to the data (top panel) and the scale dependence of the relative difference to CDM that will be potentially detectable. The middle plot shows the dependence of the power-law fits on cross-section, and the third panel shows the dependence of the power law fits on mass.}
\end{figure}

We show the projected galaxy auto-correlation of the `D' sample (Eq. \ref{eq:wgg}) in Figure \ref{fig:totalCorrelations_clustering} and the analogous IA signal (Eq. \ref{eq:wg+}) in Figure \ref{fig:totalCorrelations_alignments} for the five different dark matter models. The bottom panel of Figure \ref{fig:totalCorrelations_clustering} corresponds to the ratio between the clustering correlation in the self-interacting models and the CDM case. We note that this signal is calculated to only $1$Mpc/h. The clustering signal is suppressed at small scales relative to CDM, although this is limited to no more than $10\%$ and is only significant well-within intra-halo scales ($\lesssim 0.4$ $h^{-1}$ Mpc). Since the galaxy samples have not been cross-matched across simulations, such a change could be driven by intrinsic changes in the galaxy bias or by a selection effect.

The top panels of Figure \ref{fig:totalCorrelations_alignments} show the $w_{g,+}$ and $w_{g,\times}$ correlations of central galaxy shapes with density tracers. The left figure corresponds to SIT shapes; the right figure, to RIT shapes. We test that $w_{g,\times}$ is consistent with null through symmetry and find that the data has a $\chi^2_{\rm red}=0.8 \pm 0.4$ (SIT) and $\chi^2_{\rm red}=1.1 \pm 0.4$ (RIT) consistent with 1 for a model where $w_{g,\times}=0$ for all scales, while there is a significant negative measurement of $w_{g+}$, which indicates a projected radial alignment of S galaxies around D tracers. Moreover, RIT shapes result in a lower alignment amplitude than SIT shapes. Both results are in qualitative agreement with observational trends \citep[e.g.][]{Singh2015,SinghMandelbaum2016}. 
The bottom panels of Figure \ref{fig:totalCorrelations_alignments} show the relative difference w.r.t CDM. We find that intrinsic alignments are suppressed by dark matter interactions, with the velocity-dependent cross-section exhibiting a suppression of $w_{g+}/w_{g+,\rm CDM}>20\%$ at $r=0.1$ $h^{-1}$ Mpc. The suppression of the IA correlation can persist out to $r>1$ $h^{-1}$ Mpc from the cluster centre. The suppression can be interpreted as an overall misalignment of the central galaxy shape with the location of nearby galaxies.

\subsection{Redshift and mass dependence}
\label{sec:z} 

Intrinsic alignment correlations are known to be mass-dependent, with higher mass halos being subject to stronger alignment in line with a power-law scaling with stellar mass (or luminosity) \citep{Singh2015,vanUitert17,Piras2018}. Although not very well-constrained currently, the redshift dependence of the alignment signal is another observable which could, in the future, allow for a distinction between different alignment or dark matter models. We explore both scalings in this section. To mitigate possible differences in the alignment signal due to selection effects in the sample of central galaxies, we restrict here to the 4800 most massive halos in the simulations. (This choice corresponds to the minimum number of centrals of all redshift and dark matter models.) 

We investigate the mass dependence of the alignment signal in Figure \ref{fig:IA_mass}. We split the sample of halos into three mass bins: $13.5<\log(M/M_\odot)<14.0$; $14.0<\log(M/M_\odot)<14.5$ and $14.5<\log(M/M_\odot)<15.0$ (left to right panels) for $z=0$ and show the absolute signal in the top panels of Figure \ref{fig:IA_mass} and the signal relative to CDM in the bottom panels. We find that the alignment correlation increases with halo mass, and that the mean relative suppression inside $1$Mpc/h due to the interacting dark matter models also increases with mass, from $w_{\rm SIDM1}/w_{\rm CDM}=0.85\pm0.03$ in the lowest mass bin to $w_{\rm SIDM1}/w_{\rm CDM}=0.59\pm0.03$ in the highest mass bin. However, the highest mass bins suffers from larger variance due to the low number of clusters.

Figure \ref{fig:IA_redshift} shows the alignment signal for the four redshift bins studied. We separately analyse each of the five dark matter models (each panel), with the top panel of each showing the absolute signal and the bottom panel the signal relative to $z=0$. In each case we show increasing redshift with darker colours for the four redshifts $z=\{0.,0.125, 0.250, 0.375\}$. The top panels show that the alignment signal increases towards higher redshift for all models. This is in contrast with the clustering signal, shown in Appendix \ref{fig:redshift_clustering}, which increases towards lower redshifts (reflecting the build up of structures). This redshift dependence is similar in all models, (and between different levels of AGN heating for the CDM), which motivates us to think that this is a selection effect. However, it is over a very small redshift range and will need to be expanded to be explored completely.

\subsection{Impact of baryonic physics}

Intrinsic alignments are sensitive to the amount of baryonic feedback in hydro-dynamical simulation, representing an uncertainty in this work \citep{Marcel2014,Soussana20}. A change in the AGN heating temperature in the simulations drives changes in the alignment signal similar to those resulting from the interacting dark matter model. To distinguish between these scenarios, we study two further AGN models in the CDM case. These cases include the extreme ends of the allowed AGN heating temperature that still result in a consistent galaxy stellar mass function. They correspond, however, to a very different gas fraction in these large systems, still within observational limits \citep{BAHAMASB}. 

We measure the intrinsic alignment signals in these two simulations at two redshift slices and show the results in Figure \ref{fig:agnDependence}, where we show $z=0$ in the left and $z=0.375$ in the right hand panels, a grey shaded region denotes the upper and lower limits of the measured alignment signal, along with the five dark matter models. The bottom panel shows the suppression of the two dark matter models relative to CDM including the uncertainty in the AGN feedback. We find that at $z=0$ the AGN feedback represents a large uncertainty in the alignment signal, and without further simulations would be difficult to distinguish from SIDM. However, at higher redshifts, the uncertainty introduced by AGN is fractionally less, whilst the SIDM signal remains large. This represents a possible route to disentangling SIDM from CDM.

\subsection{Impact of shape measurement method}

\label{sec:30} Throughout our study we have used the shape of the central galaxy measured at $100$ kpc. However, the choice of shape measurement method may probe different regions of the galaxy and hence result in a different IA signal. This was already demonstrated by the differences observed in the alignment signal in the case of SIT and RIT shapes in Figure \ref{fig:totalCorrelations_alignments}. 

Here, we analyse the IA signal at a second radial scale of the central galaxy. Figure \ref{fig:scaleDep} shows the estimated signal for galaxies measured at the fiducial $r=100$ kpc and at a (conservative) smallest trusted radii of $r=30$ kpc (corresponding to a conservative five times the Plummer-equivalent softening scale of $\epsilon=4$ $h^{-1}$ kpc). Although at this scale we again observe a suppression in the IA signal, it is significantly stronger than at $100$ kpc. However, we also find that the signal is more sensitive to the choice of AGN model at this scale, and therefore does not necessarily represent an increase in the signal-to-noise.

Nevertheless, the bottom panel shows that the impact of the choice of radial scale is higher in models with increased self-interactions. As a consequence, combining two measurements of galaxy shapes at different scales could yield stronger constraints on the interaction cross-section.

\subsection{Power-law fits}

Having measured the intrinsic alignment signal in different self-interacting dark matter cosmologies, we now ask the question of whether it is possible to differentiate between CDM and SIDM in data. Given the uncertainties between observations and simulations, a scale-{\it independent} change in the intrinsic alignment signal (i.e. a shift in the amplitude) will be difficult to measure, however, a scale-dependent shift will provide a much clearer and more discriminate test. 

In order to measure the scale-dependent and independent shifts in the signal we fit an empirical power-law model to the measured IA signals: 
\be
\log_{10}\left(\frac{w_{g,+}}{{\rm Mpc}/h}\right) = \log_{10}\left(\frac{A}{{\rm Mpc}/h}\right) + \beta\left(\frac{r_p}{{\rm Mpc}/h}\right).
\ee
where $A$ is the amplitude and $\beta$, the power-law index. 
It has been shown that this is a good description of multiple small-scale alignment statistics \citep{Singh2015,Georgiou2019a,Georgiou2019b,Fortuna}. Any significant dependence of $\beta$ on the cross-section will be evidence for a scale dependent shift. 

We fit this empirical law to each observation down to a given scale radius. We define this radius by the radius at which the mean reduced chi-squared for the fit with respect to the data is closest to 1. We find that the best-fitting scale is at $r=0.23$Mpc/h. 
We show the results for the parameter fits in Figure \ref{fig:IAamplitude}. The left hand figure shows the direct fit to data of the power law and the relative difference w.r.t CDM in the bottom panel. We show the scale dependence of the relative difference for each cross-section in the solid line. The second panel shows the derived parameters from the fits as a function of cross-section with the top panel showing the shift in the amplitude (scale-independent) and the bottom panel the shift in the slope (scale-dependent). We then fit a linear relation with the interaction cross-section and find 
\be
A_{\rm 30kpc}  =  (4.56 \pm 0.17)  - (2.14 \pm 0.33) \left(\frac{\sigma_{\rm DM}/m}{{\rm cm}^2/{\rm g}}\right),
\ee
\be
A_{\rm 100kpc}  =  (6.37\pm 0.05) - (0.84 \pm 0.09) \left(\frac{\sigma_{\rm DM}/m}{{\rm cm}^2/{\rm g}}\right),
\ee
\be
\beta_{\rm 30kpc}  =  (-1.25 \pm 0.02) + (0.03 \pm 0.03) \left(\frac{\sigma_{\rm DM}/m}{{\rm cm}^2/{\rm g}}\right),
\ee
\be
\beta_{\rm 100kpc}  =  (-1.39 \pm 0.02) + (0.13 \pm 0.04 ) \left(\frac{\sigma_{\rm DM}/m}{{\rm cm}^2/{\rm g}}\right).
\ee
We find that there is a strong, scale-independent shift with cross-section, and when the shape of a galaxy is measured out to $100$kpc then we find evidence for a scale-dependent shift at 3$\sigma$confidence. This provides promising evidence that intrinsic alignments can be used to constrain self-interacting dark matter.

We study the mass dependence of each relation since we know that in the CDM cases IA is dependent on halo mass. To do this we measure the IA power-law fits and then fit a linear (in log mass) trend. The bottom panel of Figure  \ref{fig:IAamplitude} shows the mass dependence of the IA signal with the different dark matter models. We find significant evidence that an increase in halo mass shifts the amplitude of the power-law but not the slope. However, we do find that the mass dependence of the amplitude shift is clearly dependent on the cross-section, with strong self-interactions dampening the impact of increased halo mass. 

We conclude that the scale-dependent and mass dependence of the amplitude may provide a pathway to constraining SIDM with IA. For example, by constraining  a simulation calibrated $\beta$, it would be possible to compare this to observed values. Alternatively, it may be possible to normalise any observable to the high mass bin and then measure the mass dependence of the amplitude and compare that to what is expected from simulations of CDM.

\subsection{Non-linear alignment model}
Here we have fitted an empirical power-law model. However, often a more physically motivated non-linear alignment (NLA) model is used to estimate the amplitude of the IA signal \citep{Blazek2019,kidsIA}. We attempt here to fit the amplitude of the NLA to our data in the regime where it is suitable. We define `suitable' as where the reduced chi-squared of the NLA model is closest to one for the CDM model. We find this to be for scales $r_p>0.6$Mpc/$h$ ($\chi^2_{\rm red}=1.03$). We then determine the reduced chi-squared for the three other other models and find that they deviate from the NLA with increasing cross-section (for scales $r_p>0.6$Mpc/$h$). We find that the three cross-sections, $\sigma_{\rm DM}/m=0.1,0.3,1.0$ have a $\chi^2_{\rm red}=1.3,1.5,1.8$ respectively. Although SIDM deviates from the NLA model, we find that it would be difficult to use this as a method to constrain SIDM.

\section{Discussion}\label{sec:disc}
We have presented the impact of dark matter self-interactions on the intrinsic alignment of galaxies with centrals. We have found that dark matter self-interactions significantly suppress the amplitude of the intrinsic alignment signal and show how self-interactions can impact structure at the mega-parsec scale. We also present evidence for a scale-dependent shift, with self-interactions modifying the power-law index. This behaviour will provide an important feature if we are to constrain SIDM in this way. 

If we compare our results to the Horizon-AGN simulations \citep{Chisari2017} we find that the measured three-dimensional correlation function $\eta(r)$ is similar in trend to that of the satellite-satellite plus central-satellite correlation function, with the increased correlation at $\sim1$Mpc$/h$. This is interesting given that our sample contains only centrals. We hypothesise that this is due to the two-halo term of mis-identified centrals in the sample which are in fact satellites and that are large enough to enter the sample. Interestingly, \cite{Chisari2016} found that the projected intrinsic alignment signal decreased with redshift for a luminosity limited sample, whereas here we find the opposite trend, albeit at lower redshift. On the contrary, \citet{Tenneti15} found weak evolution of the projected alignment signal with redshift (in he range $0<z<1$) for a mass limited sample in the MassiveBlack-II simulation. In general, a direct comparison of BAHAMAS-SIDM with these cosmological numerical simulations is difficult given the differences in halo mass and resolution. Indeed, we carried out a verification on the clustering signal to check that this was growing with redshift and find this agrees with the literature. This dependence should be examined in future to see if it is real. However, independent of whether the CDM signal increases or decreases with redshift, we would expect the observed {\it relative} suppression due to SIDM to be larger at lower redshift since the rate of self-interactions is constant in time \citep{scatteringRates}. 

We note that the AGN and stellar feedback parameters for the BAHAMAS-SIDM simulations were not re-calibrated from the fiducial values used with CDM. This was because the changes to the calibration metrics (such as the stellar mass function) were smaller than the errors on the observational data that was being calibrated to. However, we find that the number of central galaxies that are above required threshold to get a robust shape is dependent on the cross-section of dark matter. For example the three CDM runs ( fiducial, high and low AGN reheating) have  $23,339$, $21,689$ and $24,927$ centrals respectively. The four SIDM runs, SIDM0.1, SIDM0.3, SIDM1 and vdSIDM have $22,764$, $22,784$, $22,793$ and $22,883$ respectively. We see that the number of galaxies in each SIDM run lies within the range of galaxies for the three AGN runs. As such the simulations do not require re-calibration. However, in the future as the stellar mass function becomes better constrained, hydro-dynamical SIDM simulations with the objective of precise cosmology may need to be re-calibrated in order to be consistent with observations. Moreover, we have not explored any cosmological degeneracy and whether altering the cosmological parameters can mimic the impact of self-interactions. This is something that would need to be explored in future work.

\section{Conclusions}\label{sec:conclusions}
We have used a modified version of BAHAMAS, BAHAMAS-SIDM to measure the impact of dark matter self-interactions of the intrinsic alignment of galaxies. We correlate all galaxies with the position and shape of central galaxies and measure the three-dimensional alignment and the alignment from the projected two-dimensional shape. We find that
\begin{itemize}
    \item Dark matter self-interactions induce a suppression of the alignment signal of central galaxies. This can be as high as $50\%$ depending on redshift, mass range and dark matter model.
    \item Degeneracies with AGN feedback can complicate interpretation of SIDM for low cross-sections, however we find that the range of AGN heating that is compatible with observations is distinguishable from $\sigma_{\rm DM}=0.1$\cmg.
    \item We fit empirical power-law models down to $r_{\rm cut}>0.23$Mpc/h, where the $\chi^2_{\rm red}$ is closest to one.  We find a strong scaling of the alignment amplitude and power-law index with the SIDM cross-section. However, the mass only impacts the amplitude and not the power-law index, suggesting a possible route to disentangling the CDM signal from the SIDM one.
\end{itemize}

To summarise, we have shown that self-interactions suppresses the intrinsic alignment signal of galaxies with even small interactions impacting the signal out to $\sim$Mpc scales. We show that the scale dependence of the suppression is potentially detectable with upcoming surveys such as {\it Euclid} \citep{EUCLID} or the Vera C. Rubin Observatory \citep{Ivezic}. However, to determine the redshift dependence and to categorise the scale dependence more precisely, larger simulations are required in order to garner better statistics, plus more observationally matched products such as colours and magnitudes are important if these are to be compared to data. However, it is clear that large-scale surveys can have a part to play in the quest to unveil the mystery of dark matter via intrinsic alignments.

\section*{Acknowledgements}

 We thank Harry Johnston and Cora Dvorkin for feedback that helped improve this manuscript.
 This work is part of the Delta ITP consortium, a program of the Netherlands Organisation for Scientific Research (NWO) that is funded by the Dutch Ministry of Education, Culture and Science (OCW). AR is supported by the European Research Council's Horizon2020 project `EWC' (award AMD-776247-6).

\label{lastpage}
\bibliographystyle{mn2e}
\bibliography{bibliography}

\appendix

\section{Redshift evolution of galaxy clustering}

Our results from Section \ref{sec:z} suggest that the alignment signal, $w_{g+}$, of the most massive 4800 central galaxies in our simulations increases with redshift up to $z=0.375$. In this section, we verify that the clustering of those galaxies {\it decreases} with increasing redshift, as expected for a sample where the number of galaxies is preserved across redshift bins. 

The results are shown in Figure \ref{fig:redshift_clustering}, where each panel from left to right corresponds to a different dark matter model. The bottom panels show the ratio of the clustering signal at a given redshift compared to the $z=0$ case. Darker curves indicate higher redshifts. We observe that indeed the clustering of the galaxies is consistently higher at lower redshift for all models. This is expected from our selection: the $N$ most massive galaxies at $z=0$ are expected to be more highly biased than the $N$ most massive galaxies at a higher redshift. This is in contrast to the redshift evolution of the alignments of such sample, shown in Figure \ref{fig:IA_redshift}, where we observed that alignments, as measured via $w_{g+}$, increased with increasing redshift. Our results here allow us to factor out the decreasing clustering evolution and reinforce the conclusion that the alignment of our sample increases towards higher redshifts.

\figs
\includegraphics[width=\textwidth]{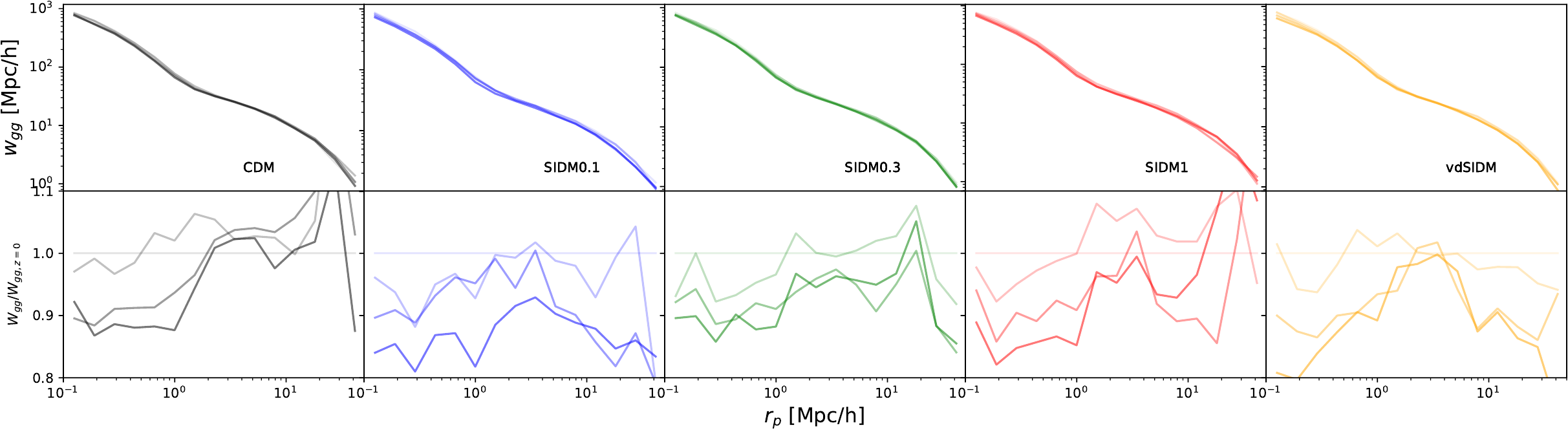}
\caption{\label{fig:redshift_clustering} The top panel shows the projected clustering statistics for the 4800 most massive galaxies in the simulations for each dark matter model (left to right). The bottom panels show the ratios of the clustering signal at a given redshift with respect to the $z=0$ case. Increasing redshifts, up to $z=0.375$, correspond to darker colours. The clustering signal is suppressed at higher redshifts for all models, as expected for a number selected sample.}
\efigs

\end{document}